\documentclass{ws-ijmpe}
\usepackage[super,compress]{cite}
\usepackage{breakurl}

\usepackage{nameref}
\usepackage{hyperref}
\hypersetup{
	colorlinks,
	citecolor=black,
	filecolor=black,
	linkcolor=black,
	urlcolor=black
}

\usepackage{graphicx}

\begin{document}

\catchline{}{}{}{}{}

\title{Classical excluded volumes of loosely bound light (anti)nuclei and their chemical freeze-out  in heavy ion collisions
%		\footnote{For the title,try not to use more than 3 lines. Typeset the title in 10 pt Times roman, uppercase and boldface.}
}

\author{Boris E.  Grinyuk}
\address{Bogolyubov Institute for Theoretical Physics, Metrologichna str. 14B, Kyiv 03680, Ukraine}

\author{Kyrill  A. Bugaev}
\address{Bogolyubov Institute for Theoretical Physics, Metrologichna str. 14B, Kyiv 03680, Ukraine\\
%%\address{
Department of Physics, Taras Shevchenko National University of Kyiv, 03022 Kiev, Ukraine\\
bugaev@th.physik.uni-frankfurt.de}

\author{Violetta V. Sagun}
\address{CFisUC, Department of Physics, University of Coimbra, Coimbra, 3004-516 Portugal\\
%%\address{
Bogolyubov Institute for Theoretical Physics, Metrologichna str. 14B, Kyiv 03680, Ukraine}
\author{Oleksii I. Ivanytskyi}
\address{CFisUC, Department of Physics, University of Coimbra, Coimbra, 3004-516 Portugal\\
%%\address{
Bogolyubov Institute for Theoretical Physics, Metrologichna str. 14B, Kyiv 03680, Ukraine}

\author{Dmitry L. Borisyuk}
\address{Bogolyubov Institute for Theoretical Physics, Metrologichna str. 14B, Kyiv 03680, Ukraine}

\author{Anatoly  S. Zhokhin}
\address{Bogolyubov Institute for Theoretical Physics, Metrologichna str. 14B, Kyiv 03680, Ukraine}

\author{Gennady M. Zinovjev}
\address{Bogolyubov Institute for Theoretical Physics, Metrologichna str. 14B, Kyiv 03680, Ukraine}

\author{David B. Blaschke}
\address{Institute of Theoretical Physics, University of Wroclaw, Max Born Pl. 1. 50-204 Wroclaw, Poland\\
%%\address{
Bogoliubov Laboratory of Theoretical Physics, JINR Dubna, Joliot-Curie Str. 6, 141980 Dubna, Russia\\
%%\address{
National Research Nuclear University (MEPhI), Kashirskoe Shosse 31, 115409 Moscow, Russia}

\author{Larissa V. Bravina}
\address{University of Oslo, POB 1048 Blindern, N-0316 Oslo, Norway}

\author{Evgeny E.  Zabrodin}
\address{University of Oslo, POB 1048 Blindern, N-0316 Oslo, Norway\\
%%\address{
Skobeltsyn Institute of Nuclear Physics, Moscow State University, 119899 Moscow, Russia}

\author{Edward G. Nikonov}
\address{Laboratory for Information Technologies,  JINR Dubna, Joliot-Curie str. 6, 141980 Dubna, Russia}

\author{Glennys Farrar}
\address{New York University, 726 Broadway, New York, New York, USA}

\author{Sonia Kabana}
\address{ Instituto de Alta Investigaci$\acute{\rm o}$n, Universidad de Tarapac$\acute{\rm a}$, Casilla 7D, Arica, Chile}

\author{Sergey V. Kuleshov}
\address{Departamento de Ciencias F$\acute{\rm i}$sicas, Universidad Andres Bello, Sazi$\acute{\rm e}$ 2212, Piso 7, Santiago, Chile}

\author{Arkadiy  V. Taranenko}
\address{National Research Nuclear University (MEPhI), Kashirskoe Shosse 31, 115409 Moscow, Russia}

\maketitle

\begin{history}
\received{Day Month Year}
\revised{Day Month Year}
%\accepted{Day Month Year}
%\comby{(xxxxxxxxxx)}
\end{history}

\begin{abstract}
From the analysis of light (anti)nuclei multiplicities  that were measured recently by the ALICE collaboration in Pb+Pb collisions at the center-of-mass collision energy $\sqrt{s_{NN}} =2.76$ TeV  there arose a highly non-trivial question about the excluded volume of composite particles. Surprisingly, the hadron resonance gas model (HRGM) is able to perfectly describe the  light (anti)nuclei multiplicities  under various assumptions. Thus, one can consider the (anti)nuclei with a vanishing hard-core radius (as the point-like particles) or with the hard-core radius of proton, but the fit quality is the same for these assumptions. It is clear, however,  that such assumptions are unphysical.   Hence we obtain  a  formula for the classical excluded volume  of  loosely bound light nuclei consisting of A baryons. To implement a new formula into the HRGM we 
have to modify  the induced surface tension concept to treat the hadrons and (anti)nuclei on the same footing.
We perform a simultaneous analysis of hadronic and (anti)nuclei multiplicities measured by the ALICE collaboration. The HRGM with the induced surface tension allows us to verify different assumptions on the values of  hard-core radii and different scenarios of chemical freeze-out of light (anti)nuclei. It is shown that the 
unprecedentedly high quality of fit  $\chi^2_{tot}/dof \simeq 0.769$  is achieved, if the chemical freeze-out temperature of hadrons  is about $T_h=150$ MeV, while the one for all (anti)nuclei is $T_A=174-175.2$ MeV. 
\end{abstract}


\ccode{PACS numbers: 25.75.Ag, 24.10.Pa}

\section{Introduction}

After almost four decades of experiments on heavy ion collisions (HIC)  which are aimed at the investigation of the 
phase diagram of quantum chromodynamics (QCD) it seems that only the low density phase of QCD known as the hadron
matter is well understood. However, a  few  puzzles  emerged recently from  the measurements of   light (anti)nuclei yields  made  by the ALICE collaboration in Pb+Pb collisions at the center-of-mass collision energy $\sqrt{s_{NN}} =2.76$ TeV \cite{KAB_Ref1a,KAB_Ref1b,KAB_Ref1c}.  The very fact that the light (anti)nuclei with binding energies of an order of a few MeV are produced at all in such violent collisions is very surprising.  These data and their analysis performed  within the simplest versions  of the hadron resonance gas model (HRGM) \cite{KAB_Jean,KAB_Ref2}  faced HIC community with  the following principal questions \cite{KAB_Jean,KAB_Ref2,PBM19}:\\
 (i) What is the mechanism of production
of  deuterons (d),  helium-3 ($^3$He),  helium-4 ($^4$He) and hyper-triton ($^3_\Lambda$H) and their antiparticles in relativistic HIC?\\
(ii) What is the thermalization mechanism of such nuclear clusters?\\
(iii)  At what temperature does  their  chemical freeze-out (CFO) occurs?

Apparently these questions are closely related to each other, but to answer them one has to correctly answer the third of them first. The problem, however, appears due to the fact that the excluded volumes of  light nuclei are not known. Therefore, the naive assumption of Ref.  \citen{KAB_Ref2} that the hard-core radius of nuclei coincides with the proton one is absolutely unphysical. A  more elaborate version of the HRGM \cite{KAB_Ref3} based on the concept of induced surface tension 
\cite{IST1,IST2}  demonstrated that  the description of light (anti)nuclei can be achieved, if their CFO temperature 
is  about 170 MeV, i.e. it is about 10\% higher than the CFO temperature  of hadrons \cite{IST1,IST2}.  

However, the results of Ref. \citen{KAB_Ref3}  were obtained using an approximate expression for the hard-core radius 
of light (anti)nuclei. Therefore, here we derive the correct expression for the classical excluded volumes of light (anti)nuclei and hyper-nuclei,
modify the HRGM with induced surface tension 
 and apply it to an analysis of the CFO temperature of such nuclei measured by  the  ALICE Collaboration \cite{KAB_Ref1a,KAB_Ref1b,KAB_Ref1c}.

The work is organized as follows. 
In Sect. 2 we remind the main ingredients of the HRGM based on the induced surface tension concept 
and briefly discuss the recent results obtained with HRGM with multicomponent hard-core repulsion among 
hadrons.
%%% the classical second virial coefficient between nuclear clusters and hadrons.
Sect. 3 is devoted to a heuristic derivation of the new EoS which treats  the hadrons and nuclear clusters
on the same footing. In Sect. 4. we present our analysis of the ALICE data \cite{KAB_Ref1a,KAB_Ref1b,KAB_Ref1c} on yields of nuclear clusters.
Our conclusions are summarized in Sect. 5.

\section{Hadron Resonance Gas Model with Induced Surface Tension}
\label{KAB_sect_HRGM}

The most advanced and successful version of the HRGM with induced surface tension \cite{IST1,IST2} 
employs   the Boltzmann statistics for all hadrons and (anti)nuclei.
Note, that such an approximation is well justified at the CFO temperatures above 50-60 MeV and it  essentially accelerates  the fitting process.
In the standard variables of  grand canonical ensemble the  HRGM  with induced surface tension  is   a system of two coupled  equations for the pressure $p$ and for the  induced surface tension  coefficient  $\Sigma$: 
\begin{eqnarray}
\label{KAB_Eq1}
&&p = \sum_{k=1}^N  p_k =
  T \sum_{k=1}^N \phi_k \exp \left[ \frac{\mu_k - V_k p - S_k \Sigma}{T} \right]
\,, ~\\
\label{KAB_Eq2}
&&\Sigma =
\sum_{k=1}^N  \Sigma_k = 
 T \sum_{k=1}^N R_k \phi_k \exp 
\left[ \frac{\mu_k - V_k p -  S_k\alpha \Sigma}{T} \right] \,.
\end{eqnarray}
Here $p_k$ and $\Sigma_k$ are, respectively, the partial pressure and partial induced surface tension coefficient of the
$k$-th sort of particle, while  
  $V_k = \frac{4}{3}\pi R_k^3$ denotes its proper volume, $S_k = 4\pi R_k^2 $ is its    proper surface and
  $R_k$ is its hard-core radius. The thermal density of particle
  of mass $m_k$, degeneracy $g_k$ and chemical potential $\mu_k$ is $ \phi_k (T) =  g_k\int  \frac{dp^3}{(2\pi\hbar)^3} \exp\left[- \frac{\sqrt{p^2+m_k^2}}{T} \right]$.
  
  The coefficient  $\alpha = 1.25$ \cite{IST1,IST2} allows one to go beyond the 
  second virial coefficient approximation and to correctly  consider the high values of packing fractions up to  
  $\eta \equiv \sum_k \frac{4}{3}\pi R_k^3 \rho_k 
\simeq  0.2-0.22$. 
It is highly nontrivial, that a single parameter $\alpha$ allows one to correctly reproduce the third and the fourth
virial coefficient of the classical hard spheres \cite{IST1,IST2, Ivanytskyi18,QIST18,QIST19}.
Therefore, such a version of HRGM allows one to access  the  high  values of  particle number 
densities for which the  HRGM  based on the usual Van der Waals  approximation \cite{PBM06} is entirely wrong.  

Using the partial  values $p_k$ and $\Sigma_k$ one can write the particle number density of $k$-th sort of particle as
\begin{equation}\label{KAB_Eq3}
\rho_k \equiv \frac{\partial  p}{\partial \mu_k} = \frac{1}{T} \cdot \frac{p_k \, a_{22} 
- \Sigma_k \, a_{12}}{a_{11}\, a_{22} - a_{12}\, a_{21} } \,,
\end{equation}
where the coefficients $a_{kl}$ are defined as
\begin{eqnarray}\label{KAB_Eq4}
&& a_{11} =  1 + \frac{4}{3} \pi \sum_k  R_k^3 \frac{p_k}{T} \,, \quad 
a_{12}  =  4 \pi \sum_k R_k^2 \frac{p_k}{T} \, , \\
\label{KAB_Eq5}
&&a_{21} = \frac{4}{3} \pi \sum_k R_k^3\frac{\Sigma_k}{T} \, , \quad
a_{22} = 1 + 4 \pi \sum_k  R_k^2 \alpha \frac{\Sigma_k}{T} \,.
\end{eqnarray}
Then using Eqs. (\ref{KAB_Eq3})-(\ref{KAB_Eq5}) one can find the absolute thermal yield $N_k^{th} = V \rho_k$ of particles and their  thermal ratios ${\cal R}^{th}_{kl} = \frac{N_k^{th}}{N_l^{th}}$. 
For the known  branchings  $Br_{l\rightarrow k}$ of hadronic  decays $l\rightarrow k$  one can 
determine the total yield of $k$-th sort of  hadrons as
\begin{eqnarray}
\label{KAB_Eq6}
N^{tot}_k = V\left[ \rho_k+\sum_{l\neq k}\rho_l\, Br_{l\rightarrow k} \right] \,,
\end{eqnarray}
where $V$ is the CFO volume.  
Since all details of the  fitting process   are  well documented, we refer to the original works \cite{IST1,IST2}.

Another great advantage of the induced surface tension approach is that the number of equations for pressure $p$ and induced surface tension coefficient  $\Sigma$ is always two and it does not depend on the number of different hard-core radii. This advantage  essentially reduces the CPU time of the fitting process, in the number of different hard-core radii is larger than 3. 

It is necessary to stress that the high quality of description of  the experimental multiplicities measured at AGS  BNL, SPS CERN, RHIC BNL  and LHC CERN accelerators was mainly  achieved by the HRGM with the multicomponent hard-core repulsion\cite{IST1,IST2, IST3,HRGM13a,Sagun14}, i.e. with several hard-core particles of hadrons.  After the different hard-core radii of pions $R_\pi$, kaons $R_K$, of other mesons $R_m$ and baryons $R_b$ were implemented into the HRGM and used to describe the experimental hadronic multiplicities in Ref.  \citen{HRGM13a},  the   HRGM  became a reliable and extremely successful tool of 
HIC  phenomenology to extract the parameters of chemical freeze-out of  high energy nuclear collisions, i.e. the moment after which the inelastic reactions stop to exist, while the hadronic matter  evolution is governed by the  elastic reactions and the decays of resonances \cite{PBM06}. The HRGM  with multicomponent repulsion not only allowed
us to achieve the highest quality of the hadronic multiplicity description with the  smallest values of $\chi^2/dof$ being in the range between 0.96 \cite{Sagun14} and 1.1 \cite{IST2},  but it also allowed us for the first time to simultaneously  describe the  
peaks in the collision energy dependence of $K^+/\pi^+$ and $\Lambda/\pi^-$ ratios  without spoiling the other hadronic ratios 
\cite{HRGM13a}. Moreover, the HRGM  with multicomponent repulsion allowed us to demonstrate that the 
$\gamma_s$-factor introduced in Ref. \citen{Rafelski91} to quantify the deviation of strange charge from chemical equilibrium is not necessary to describe the hadronic multiplicities, since they can be perfectly  explained by the fact that the CFO of  strange hadrons occurs on a separate hyper-surface compared to the hadrons consisting of $u$ and $d$ (anti)quarks \cite{SFO}.  Note that the hypothesis of separate CFO of strange hadrons can be justified even, if the hadrons are treated as an ideal gas \cite{SFO_ind}. More sophisticated scenarios of the CFO of strange particles  can be found in Refs. \citen{SFO_ind2, SFO2}. 

Furthermore, the successful  description of the experimental hadronic multiplicities   allowed us to  uncover several irregularities in the energy dependencies of various thermodynamic and hydrodynamic quantities at CFO
which were explained as possible  experimental signals of two QCD phase transitions   \cite{GSA15,GSA16,GSA16b,Signals18,Signals19}. 
The most remarkable irregularities include two sets of highly  correlated quasi-plateaus found in  \cite{GSA15,GSA16} which are located at the collision energy ranges
$\sqrt{s_{NN}} = 3.8-4.9$ GeV and $\sqrt{s_{NN}} = 7.6-9.2$ GeV, and two peaks of trace anomaly  $\delta = (\epsilon - 3p)/T^4$ observed at the maximal energy of each set of highly  quasi-plateaus (here $\epsilon$,  $p$ and $T$ denote, respectively, the energy density of the system, its pressure and temperature). Using the HRGM  with multicomponent repulsion
it was possible to find out  two strong peaks of the baryonic charge density located
at $\sqrt{s_{NN}} = 4.9$ GeV and $\sqrt{s_{NN}} = 9.2$ GeV, i.e.
 exactly at the collision energies of the trace anomaly peaks. 

The hydrodynamical signals, i.e. the highly correlated quasi-plateaus,  in the low collision energy region  were predicted a long time ago  in Refs. 
\citen{KAB:89a, KAB:90} as a manifestation of the mixed phase  with the anomalous thermodynamic properties. 
Moreover,  the generalized shock adiabat model \cite{KAB:89a, KAB:90} allowed us to 
establish  the one-to-one correspondence between the peak of trace anomaly $\delta$ at CFO and a similar peak of $\delta$  at 
the  generalized shock adiabat \cite{GSA15,GSA16} located at the boundary of the mixed and high density phases. 
Besides, at CFO the low energy signals are accompanied by the strong jumps of the pressure $p$ and the effective
number of degrees of freedom $p/T^4$  \cite{GSA15,GSA16,GSA16b}. 
The  high energy irregularities  are  less pronounced and, hence, their  interpretation was given
later \cite{Signals18,Signals19}.

Very recently the advanced versions of HRGM helped us to estimate the number of the bosonic and fermionic degrees of freedom of the nearly massless  matter that  is created in HIC at the collision energies  $\sqrt{s_{NN}} = 6.3-9.2$ GeV
which, respectively, are $N_B \simeq 1520$ and $N_F \simeq 140$  \cite{Signals18}.  This finding allowed us to interpret 
the two sets of irregularities \cite{Signals18,Signals19} as the 1-st order phase transition from normal hadronic matter to the hadronic matter with partially restored chiral symmetry in the non-strange sector at $\sqrt{s_{NN}} = 3.8-4.9$ GeV and the 2-nd order  phase transition (or a very weak one of the 1-st order) from nearly massless non-strange hadronic matter to quark gluon plasma (deconfinement) at  $\sqrt{s_{NN}} = 9-10$ GeV.  Remarkably, but a similar conclusion  with very similar threshold collision energies of  two QCD phase transitions  
was made  in Refs.  \citen{Cassing16,Cassing16b} using the most advanced transport approach known as the Parton-Hadron-String-Dynamic-Model. Moreover, it is necessary to stress  the idea  that  deconfinement phase transition is of 2-nd order 
was first suggested not by the lattice formulation of QCD, but by the QCD phenomenology in Refs. \citen{Sonia1,Sonia2}.

From this extended reminder on the recent findings based on the HRGM with realistic hard-core repulsion one can see that 
 essential improvement of the data description always led to a deep insight in the understanding of the QCD matter properties.
Therefore, one can expect that the correct theoretical approach to the CFO of nuclear clusters in high energy HIC will provide us with extremely valuable information about the properties of matter from which such clusters  are produced.

\section{Excluded volumes of light nuclear clusters}

The main problem with the  excluded volumes of light (anti)nuclei consisting of $A$ number of baryons  ($2 \le A \le 4$)  is not that they are loosely bound compared to the typical temperature $T \simeq 150-160$ MeV, but that they are the clusters themselves.
Therefore, the usual Mayer procedure to calculate their cluster integrals and the excluded volumes cannot be used.  
However, the fact that  the small nuclei of $A$ baryons are roomy clusters, i.e. their radius ${\cal R}_A \sim 1.1 (A)^\frac{1}{3}$ fm \cite{KAB_Bohr} is much  larger than the maximal double hard-core radius of hadrons $2 R_m \simeq 0.84$ fm \cite{IST1,IST2},
allows one to easily find out the desired excluded volume. Indeed, using this fact one can freely translate the hadron of 
hard-core radius $R_h$ around each of the constituent (baryon of hard-core radius $R_b$)  of a considered  nuclear cluster   without touching  any other constituents inside  this nucleus. 
Therefore, the resulting  excluded volume  (per particle) of a hadron and the nucleus of $A$ baryons can be written as
 \begin{equation}\label{KAB_Eq7}
 b_{Ah} =  A \frac{2}{3}\pi (R_b+R_h)^3\,, 
\end{equation}
where $R_b$ is  the hard-core radius of baryons.  Unfortunately, one cannot use such excluded volumes in the system 
(1)-(5) and, therefore, our first  task  is to simplify Eq. (\ref{KAB_Eq7}).

From Eq. (\ref{KAB_Eq7}) one can find the equivalent hard-core radius $R^{eq}_{Ah}$
of a pair $Ah$   by equating two excluded volumes $\frac{2}{3}\pi (R_{Ah}^{eq})^3 = b_{Ah}$. Then we get
the equivalent hard-core radius as 
 \begin{equation}\label{KAB_Eq8}
 R_{Ah}^{eq} =  A^\frac{1}{3}(R_b+ R_h)\, ,
\end{equation}
which can be used to determine the effective hard-core radius of a nucleus 
in a medium completely   dominated by pions 
 \begin{equation}\label{KAB_Eq9}
R_A \simeq R_{A\pi}^{eq} - R_\pi \simeq  A^\frac{1}{3}R_b + (A^\frac{1}{3}-1) R_\pi \simeq  A^\frac{1}{3}R_b\, .
\end{equation}
Such an approximation is well  justified by the fact that pions are  the most abundant particles at ALICE  collision energy and their hard-core radius $R_\pi \simeq 0.15$ fm  \cite{IST1,IST2} is substantially smaller than $R_b =0.365$ fm. 
Even for $A=4$ the correction $(A^\frac{1}{3}-1) R_\pi \le 0.088$ fm neglected  in Eq.  (\ref{KAB_Eq9})  is very small compared 
to the effective hard-core radius of nuclei $R_A  \simeq  A^\frac{1}{3}R_b$.
 Nevertheless, to  verify   the  validity of the HRGM (1)-(5) with the  approximate  expression  (\ref{KAB_Eq9})  
below we derive an alternative EoS  which employes 
the   true excluded volumes  (\ref{KAB_Eq7}). 

Our derivation of the alternative EoS  is based on the heuristic approach of Ref. \citen{KAB_IST0}.
In the variables of the grand canonical ensemble the  pressure of the mixture of gases of hadrons and nuclear clusters
can be written as (cluster expansion) \cite{KAB_IST0}  
\begin{eqnarray}\label{KAB_Eq10}
&& \hspace*{-2.2mm}p \simeq  T \sum\limits_{h} \tilde \phi_h + T \sum\limits_{A} \tilde \phi_A - T \sum\limits_{h}  \sum\limits_{h^\prime} \tilde \phi_h
b_{hh^\prime} \tilde  \phi_{h^\prime}  - 2 T  \sum\limits_{h}  \sum\limits_{A} \tilde \phi_h
b_{hA}\tilde  \phi_{A} . \qquad 
\end{eqnarray}
The first two terms on the right hand side of Eq. (\ref{KAB_Eq10}) represent the pressure of ideal gases of hadrons
and nuclei, while the other terms describe the hard-core repulsion between all particles.  The excluded volumes of hadrons 
are given by the expression $b_{hh^\prime} = \frac{2}{3} \pi (R_h+R_{h^\prime})^3$, while the excluded volumes $b_{hA}$  are defined  by Eq.  (\ref{KAB_Eq7}).  The last term in Eq. (\ref{KAB_Eq10}) accounts for the excluded volumes $b_{Ah}= b_{hA}$ as well,  but it is doubled to shorten the notations.  Evidently, the hard-core repulsion between two nuclear clusters can be neglected due to their low particle number density. 

To simplify the evaluation in Eq. (\ref{KAB_Eq10}) we introduced the particle number density of the ideal Boltzmann gas of sort $n$ as
	\begin{equation} \label{KAB_Eq11}
	\tilde \phi_n = g_n \int  \frac{dk^3}{(2\pi\hbar)^3} \exp\left[ \frac{\mu_n -\sqrt{p^2+m_n^2}}{T} \right] .
	\end{equation}
Substituting into Eq. (\ref{KAB_Eq10})  the expressions for the excluded volumes, one finds
\begin{eqnarray}\label{KAB_Eq12}
&& \hspace*{-2.2mm}p \simeq  T \sum\limits_{h} \tilde \phi_h + T \sum\limits_{A} \tilde \phi_A - T \sum\limits_{h}  \sum\limits_{h^\prime} \tilde \phi_h \frac{2}{3} \pi \left[R_h^3 + 3R_h^2 R_{h^\prime} +  3R_h R_{h^\prime}^2 + R_{h^\prime}^3\right] \tilde  \phi_{h^\prime} - \nonumber \\
   && \hspace*{-2.2mm}-  T  \sum\limits_{h}  \sum\limits_{A} \tilde \phi_h
   \frac{4}{3} \pi \left[R_h^3 + 3R_h^2 R_{b} +  3R_h R_{b}^2 + R_{b}^3\right]
A \tilde  \phi_{A}  \simeq  \\
 && \hspace*{-2.2mm}
\simeq  T \sum\limits_{h} \tilde \phi_h + T \sum\limits_{A} \tilde \phi_A - T \sum\limits_{h}  \tilde \phi_h \frac{4}{3} \pi  R_h^3 \left[ \sum\limits_{h^\prime} \tilde  \phi_{h^\prime}\right]- T \sum\limits_{h}  \tilde \phi_h 4 \pi  R_h^2 \left[  \sum\limits_{h^\prime} R_{h^\prime} \tilde  \phi_{h^\prime}\right] -
 \nonumber \\
   && \hspace*{-2.2mm}
 - T \sum\limits_{h}  \tilde \phi_h \frac{4}{3} \pi  R_h^3 \left[  \sum\limits_{A} A \tilde  \phi_{A}\right] - T \sum\limits_{h}  \tilde \phi_h 4 \pi  R_h^2 \left[  \sum\limits_{A} R_{b} A \tilde  \phi_{A}\right] - \nonumber \\
   && \hspace*{-2.2mm}
   -  T   \sum\limits_{A}  \tilde \phi_A \left[ 
   \frac{4}{3} \pi R_b^3 A \left[  \sum\limits_{h^\prime} \tilde  \phi_{h^\prime}\right] + 4\pi R_{b}^2 A \left[  \sum\limits_{h^\prime} R_{h^\prime} \tilde  \phi_{h^\prime}\right]\right]
 . \qquad 
 \label{KAB_Eq13}
\end{eqnarray}
Regrouping in Eq. (\ref{KAB_Eq13}) the terms with partial  pressure of ideal gas $p_i \sim T \tilde \phi_i$ and the 
partial coefficient of induced surface tension of ideal gas $\Sigma_i \sim T R_i \tilde \phi_i$, where  $R_i = \{ R_h, A R_b\}$,
one can write 
\begin{eqnarray}\label{KAB_Eq14}
&& \hspace*{-8.2mm}p \simeq  T \sum\limits_{h} \tilde \phi_h  \times \nonumber \\
&& \hspace*{-8.2mm}
 \times \left[ 1  -  \frac{4}{3} \pi  R_h^3 \left(  \sum\limits_{h^\prime} \tilde  \phi_{h^\prime}+ \sum\limits_{A} A \tilde  \phi_{A}\right) -  4 \pi  R_h^2 \left(  \sum\limits_{h^\prime} R_{h^\prime} \tilde  \phi_{h^\prime}
+  \sum\limits_{A} R_{b} A \tilde  \phi_{A}\right)
\right] +
 \nonumber \\
  && \hspace*{-8.2mm} +     T \sum\limits_{A} \tilde \phi_A  \left[ 1 
   -  
   \frac{4}{3} \pi R_b^3 A \left(  \sum\limits_{h^\prime} \tilde  \phi_{h^\prime}\right) - 4\pi R_{b}^2 A \left(  \sum\limits_{h^\prime} R_{h^\prime} \tilde  \phi_{h^\prime}\right)\right]
 . 
\end{eqnarray}
Recalling that at low particle number densities the pressure can be approximated by the ideal gas term, one can rewrite 
Eq. (\ref{KAB_Eq14}) as the following system
\begin{eqnarray}\label{KAB_Eq15}
&& \hspace*{-2.2mm}p \simeq  T \sum\limits_{h} \tilde \phi_h \left[ 1  -  \frac{4}{3} \pi  R_h^3 \left( \frac{p}{T}+ \sum\limits_{A} (A-1) \tilde  \phi_{A}\right) -  4 \pi  R_h^2 \frac{\Sigma}{T}\right]
 \nonumber \\
   && \hspace*{-2.2mm}
   +   T \sum\limits_{A} \tilde \phi_A  \left[ 1 
   -  
   \frac{4}{3} \pi R_b^3 A \left(  \frac{p}{T} + \sum\limits_{A} (A-1) \tilde  \phi_{A}  \right) - 4\pi R_{b}^2 A \frac{\Sigma}{T} \right]
 , \qquad  \\
&& \hspace*{-2.2mm} \Sigma  \simeq  T \left[  \sum\limits_{h^\prime} R_{h^\prime} \tilde  \phi_{h^\prime}
+  \sum\limits_{A} R_{b} A \tilde  \phi_{A} \right] .
\label{KAB_Eq16}
\end{eqnarray}
Apparently we can  safely neglect  the small term $T \sum\limits_{A} (A-1) \tilde  \phi_{A} $ compared to  the pressure $p$ 
in Eq. (\ref{KAB_Eq15}).
Furthermore,  the following modification of  the particle number densities $\tilde  \phi_{h}$ and $\tilde  \phi_{A} $
\begin{eqnarray}   \label{KAB_Eq17}
&&\tilde  \phi_{h} \rightarrow \tilde  \phi_{h}  \exp \left[  -  \frac{4}{3} \pi  R_h^3  \frac{p}{T} - \alpha  4 \pi  R_h^2 \frac{\Sigma}{T}\right] , \\
&&\tilde  \phi_{A}  \rightarrow \tilde    \phi_{A}  \exp \left[ 
   -   \frac{4}{3} \pi R_b^3 A   \frac{p}{T}  - \alpha 4\pi R_{b}^2 A \frac{\Sigma}{T} \right] , 
   \label{KAB_Eq18}
\end{eqnarray}
in Eq. (\ref{KAB_Eq16}) will  change only the higher order terms of cluster expansion which can be corrected later on
by proper choice of parameter $\alpha$ \cite{QIST19}. 
Taking this into account, now we can  extrapolate the system  (\ref{KAB_Eq15}),  (\ref{KAB_Eq16})
to high densities as
\begin{eqnarray}\label{KAB_Eq19}
p_v &=&   T \sum\limits_{h} \tilde \phi_h \exp \left[  -  \frac{4}{3} \pi  R_h^3 \frac{p_v}{T}  -  4 \pi  R_h^2 \frac{\Sigma_v}{T}\right]
 \nonumber \\
  &+&    T \sum\limits_{A} \tilde \phi_A \exp \left[ 
   -  
   \frac{4}{3} \pi R_b^3 A   \frac{p_v}{T} - 4\pi R_{b}^2 A \frac{\Sigma_v}{T} \right]
 , \qquad  \\
\Sigma_v  &=&  T  \sum\limits_{h} R_{h} \tilde  \phi_{h}  \exp \left[  -  \frac{4}{3} \pi  R_h^3 \frac{p_v}{T}  - \alpha  4 \pi  R_h^2 \frac{\Sigma_v}{T}\right] +
 \nonumber \\
&+& T  \sum\limits_{A} R_{b} A \tilde  \phi_{A} \exp \left[ 
   -  
   \frac{4}{3} \pi R_b^3 A  \frac{p_v}{T} - \alpha 4\pi R_{b}^2 A \frac{\Sigma_v}{T} \right] ,
   \label{KAB_Eq20}
\end{eqnarray}
where the parameter $\alpha =1.245$ is introduced to account for higher order classical virial coefficients as it is shown
in Refs. \citen{IST1,IST2,QIST19}.  The subscript $v$ in the quantities  $p_v$ and $\Sigma_v$  of the  system  (\ref{KAB_Eq19}), (\ref{KAB_Eq20}) is introduced to indicate that it is obtained from the excluded volume 
expression (\ref{KAB_Eq7}) of nuclear cluster of $A$ baryons  and a hadron $h$. 
This system 
can be obtained more rigorously 
by the Laplace transform method in a spirit of Ref. \citen{KAB_Nazar}, but such a derivation is even more complicated than the one presented above. 

The expressions (\ref{KAB_Eq3})-(\ref{KAB_Eq5}) can be used to calculate the particle number densities of nuclear clusters
for the system (\ref{KAB_Eq19}), (\ref{KAB_Eq20}),
if one makes the following replacements $R_A^2 \rightarrow A R_b^2$ and $R_A^3 \rightarrow A R_b^3$ in Eqs. 
 (\ref{KAB_Eq4})-(\ref{KAB_Eq5}).

Rewriting  the systems (1), (2) in terms of the effective radius of the nuclear cluster (\ref{KAB_Eq9}), one obtains
\begin{eqnarray}\label{KAB_Eq21}
p_r &=&   T \sum\limits_{h} \tilde \phi_h \exp \left[  -  \frac{4}{3} \pi  R_h^3 \frac{p_r}{T}  -  4 \pi  R_h^2 \frac{\Sigma_r}{T}\right]
 \nonumber \\
  &+&    T \sum\limits_{A} \tilde \phi_A \exp \left[ 
   -  
   \frac{4}{3} \pi R_b^3 A   \frac{p_r}{T} - 4\pi R_{b}^2 A^\frac{2}{3} \frac{\Sigma_r}{T} \right]
 , \qquad  \\
\Sigma_r  &=&  T  \sum\limits_{h} R_{h} \tilde  \phi_{h}  \exp \left[  -  \frac{4}{3} \pi  R_h^3 \frac{p_r}{T}  - \alpha  4 \pi  R_h^2 \frac{\Sigma_r}{T}\right] +
 \nonumber \\
&+& T  \sum\limits_{A} R_{b} A^\frac{1}{3} \tilde  \phi_{A} \exp \left[ 
   -  
   \frac{4}{3} \pi R_b^3 A  \frac{p_r}{T} - \alpha 4\pi R_{b}^2 A^\frac{2}{3} \frac{\Sigma_r}{T} \right] ,
   \label{KAB_Eq22}
\end{eqnarray}
where we introduced the subscript $r$ in the functions $p_r$ and $\Sigma_r $ to signify the fact that it is obtained 
using an effective hard-core radius of the nuclear cluster  (\ref{KAB_Eq9}). 

Comparing the systems (\ref{KAB_Eq19}), (\ref{KAB_Eq20}) and (\ref{KAB_Eq21}), (\ref{KAB_Eq22}), one can see 
that only the eigen volumes of nuclear clusters are the same, while the eigen surface $4\pi R_{b}^2 A$ and the eigen radius $R_{b} A$ of a nucleus of $A$ baryons of the system (\ref{KAB_Eq19}), (\ref{KAB_Eq20}) are larger than the corresponding quantities of the system (\ref{KAB_Eq21}), (\ref{KAB_Eq22}). Therefore, for the same values of $T$ and $\{\mu_k\}$ the partial pressure of such nuclear cluster and, consequently, its particle number density,  will be smaller for the system  (\ref{KAB_Eq19}), (\ref{KAB_Eq20}). In other words, to obtain the same particle number density of  nuclear clusters  of $A$ baryons the temperature or chemical potentials of  the system 
(\ref{KAB_Eq19}), (\ref{KAB_Eq20}) should be larger then for the system (\ref{KAB_Eq21}), (\ref{KAB_Eq22}).  
Nevertheless, since the system (\ref{KAB_Eq21}), (\ref{KAB_Eq22})  was obtained under the reasonable approximation,
then for the pion dominated hadronic matter  the results obtained from the EoS  (\ref{KAB_Eq21}), (\ref{KAB_Eq22}) should 
be very similar to the ones found from the EoS (\ref{KAB_Eq19}), (\ref{KAB_Eq20}). This will be our guideline for the analysis of ALICE data \cite{KAB_Ref1a,KAB_Ref1b,KAB_Ref1c}.

\section{Details of fitting procedure}

\begin{figure}[th]
	\centerline{\includegraphics[width=77mm]{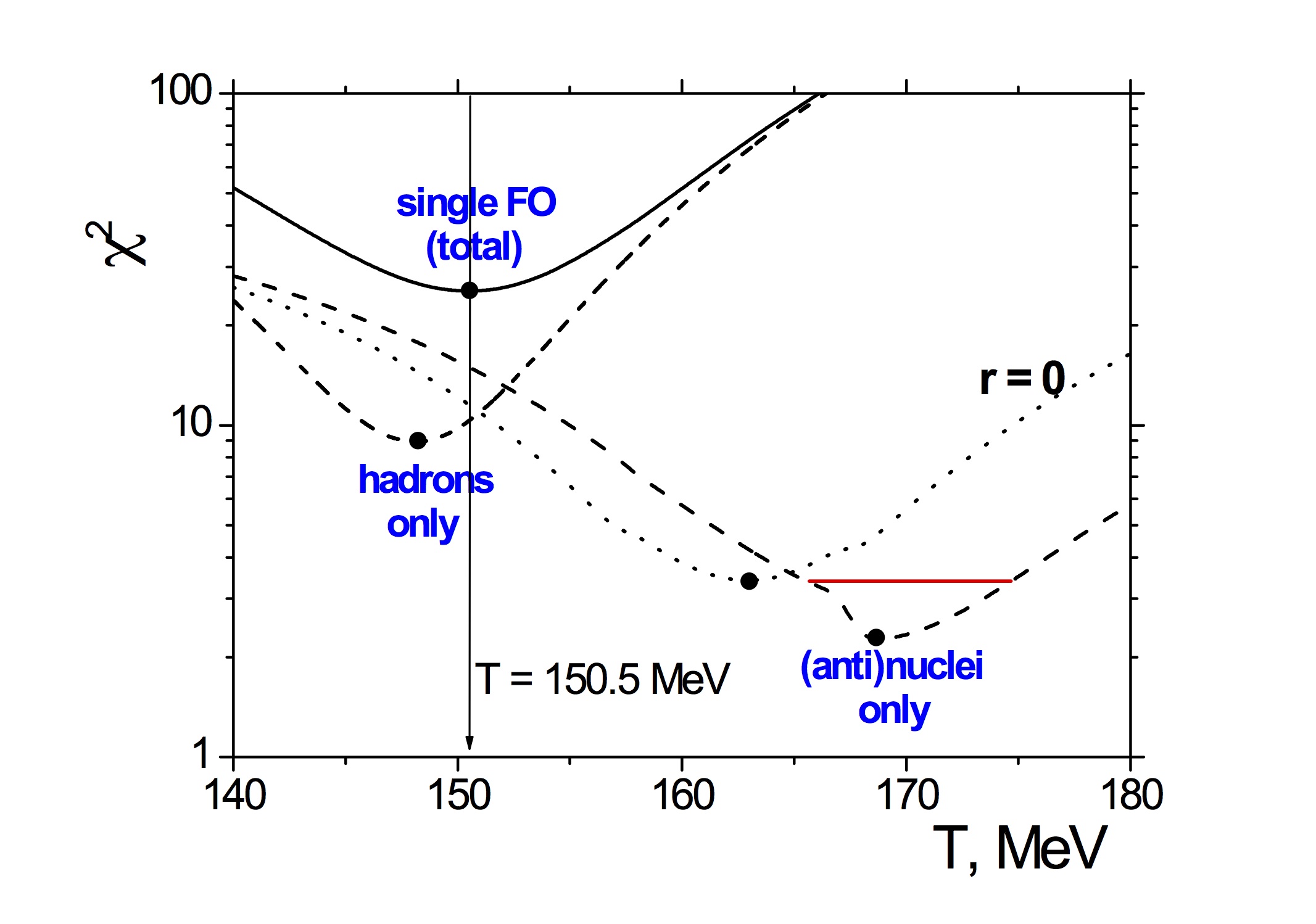}
	\hspace*{-11mm}
	\includegraphics[width=77mm]{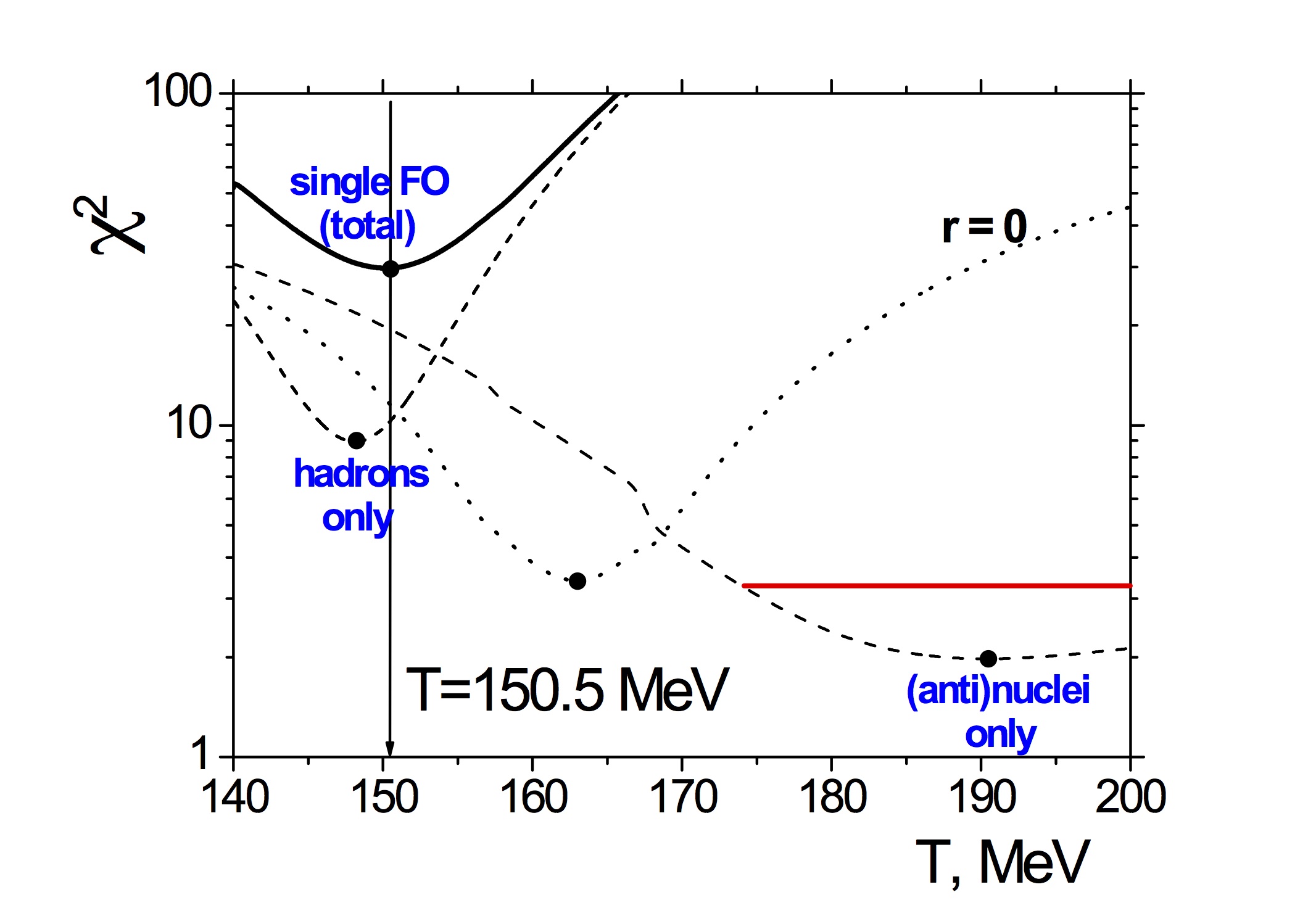}
	}

	\caption{ {\bf Left panel.} $\chi^2$ as a function of  the CFO temperature $T$ for the ALICE data: hadrons only (short dash curve), nuclear clusters only for the EoS (\ref{KAB_Eq21}), (\ref{KAB_Eq22})  (short dashed curve)  and their sum 
	 (solid curve). For a comparison the $\chi^2_A$ of  nuclear clusters  with the vanishing size is shown by the dotted curve.  Horizontal bar on top of the $\chi^2$  minimum of nuclear clusters  shows the deviations that correspond to 
 $\min(\chi^2) +1$ .     {\bf Right panel.} Same as in the left panel, but for the EoS
(\ref{KAB_Eq19}), (\ref{KAB_Eq20}).
}
	\label{KAB_Fig1}
\end{figure}

Two formulations of the HRGM with induced surface tension, i.e. R- and V-approaches,  described in the preceding section are applied to describe the whole set of the ALICE data measured at $\sqrt{s_{NN}} =2.76$ TeV.
Since the hadronic multiplicities are very well described by the HRGM with the induced surface tension with $\chi^2_h/dof \simeq 10.7/(11-1) \simeq 1.07$  \cite{IST2} here  we do not discuss the results of this analysis. 
The most important assumption  used to analyze the ALICE data measured at $\sqrt{s_{NN}} =2.76$ TeV is that all chemical potentials are set to zero \cite{IST1,IST2}.
More  details  describing the  
HRGM, the analyzed  data and the  fitting process can be found  in Refs. \citen{IST1,IST2}.
Hence, here our  main attention is payed  to the fitting of the nuclear clusters data \cite{KAB_Ref1a,KAB_Ref1b,KAB_Ref1c} for the centrality $0-10$\%.  

First we discuss the traditional approach in which 
 it is  assumed that  the CFO  occurs for  all particles simultaneously, i.e. hadrons and (anti)nuclei, are freezing out from  a single hyper-surface of CFO.
Then the total $\chi^2_{tot}(V)$ of such a model (M1 hereafter) is as follows 
\begin{eqnarray}\label{KAB_Eq23}
\hspace*{-2.2mm}\chi^2_{tot}(V) &=& \chi^2_{h} + \chi^2_{A} (V)
= \sum_{{\rm pairs}\,\, {k,l}} \left[  \frac{{\cal R}_{kl}^{theo} - {\cal R}_{kl}^{exp}}{\delta {\cal R}_{kl}^{exp}}\right]^2  +   \sum_A \left[  \frac{\rho_A(T) V - N^{exp}_A}{\delta N^{exp}_A}\right]^2  . \quad
\end{eqnarray}
Here $\chi^2_{h}$ and $\chi^2_{A}$ denote, respectively,  the properly normalized  mean deviation squared for hadrons and (anti)nuclei. 
According to Refs. \citen{IST1,IST2} the hadronic part of $\chi^2_{tot}(V)$ contains  only the ratios of hadronic  multiplicities ${\cal R}_{kl}$, since in this case one  does not need the CFO   volume for fitting.  Such a fitting procedure  has several advantages 
and was successfully  used in our previous publications \cite{IST1,IST2,HRGM13a,Sagun14,SFO,SFO2,Signals18,Signals19}.
It, however, is inconsistent with the   (anti)nuclei part of $\chi^2_{tot}(V)$  which depends on the thermal densities of (anti)nuclei of $A$ (anti)baryons and the CFO  volume $V$. 
The simplest way  to determine  the CFO volume  $V$ is to 
%%%relate it to   the multiplicity of $\pi^+$-mesons as
%%$V = \frac{N^{exp}_{\pi^+}}{\rho_{\pi^+}}$.  
use the maximum likelihood method, i.e. for each value of CFO temperature $T$ to find the minimum of $\chi^2_{tot}(V)$ 
with respect to the CFO volume  $V$ and to determine the function $V(T)$ from the equation $
\frac{\partial \chi^2_{tot}(V)}{\partial  V} =0$.
Then  the model has two fitting parameters, namely the CFO temperature and volume.
We verified  that  another way of fitting by introducing the ratios of (anti)nuclei to the multiplicity of 
$\pi^+$-mesons, like it is done for the hadronic ratios, 
provides practically  the same result for the CFO temperature, but, unfortunately, the resulting error for the ratio of multiplicities  is larger and, hence, the resulting $\chi^2_{tot}$ is slightly smaller.

As one can see from Fig. \ref{KAB_Fig1}  the minimum of $\chi^2_{tot}$ for the M1 is achieved at the same CFO temperature 
$T_{M1} \simeq 150.5 \pm 6$ MeV for both EoS. This is a consequence  of the fact that the main contribution to $\chi^2_{tot}$
is defined by the hadrons, since it strongly increases with increasing $T_{M1}$. This can also be  seen from the both panels of  Fig. \ref{KAB_Fig1}. However, the model M1R defined by the EoS (\ref{KAB_Eq21}), (\ref{KAB_Eq22})  provides slightly better value of  $\chi^2_{tot}/dof|_{M1R} \simeq
(10.1+15)/(11+8-2)  = 25.1/17 \simeq 1.476$, where 11 is the number of  fitted hadronic ratios, 8 is the number of analyzed 
data point for the nuclear clusters. For the model M1V defined by the EoS (\ref{KAB_Eq19}), (\ref{KAB_Eq20}) 
one finds that $\chi^2_{tot}/dof|_{M1V} \simeq (10.1+21)/(11+8-2)  = 31.1/17 \simeq 1.83$.  Thus, we got  a very strange 
and surprising result that the more elaborate model describes the same set of data somewhat worse than the approximative one. 

\begin{figure}[t]
	
	\centerline{\includegraphics[width=63mm]{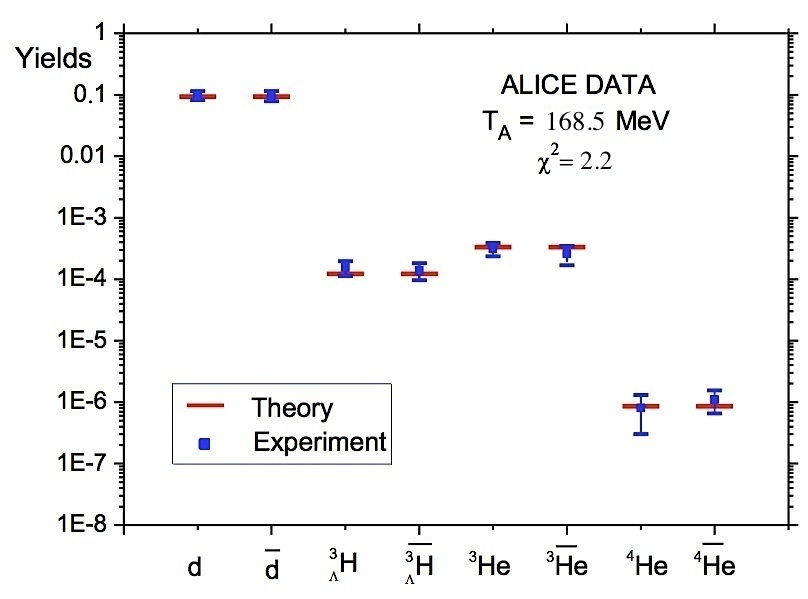}
	\hspace*{-2mm}
	\includegraphics[width=63mm]{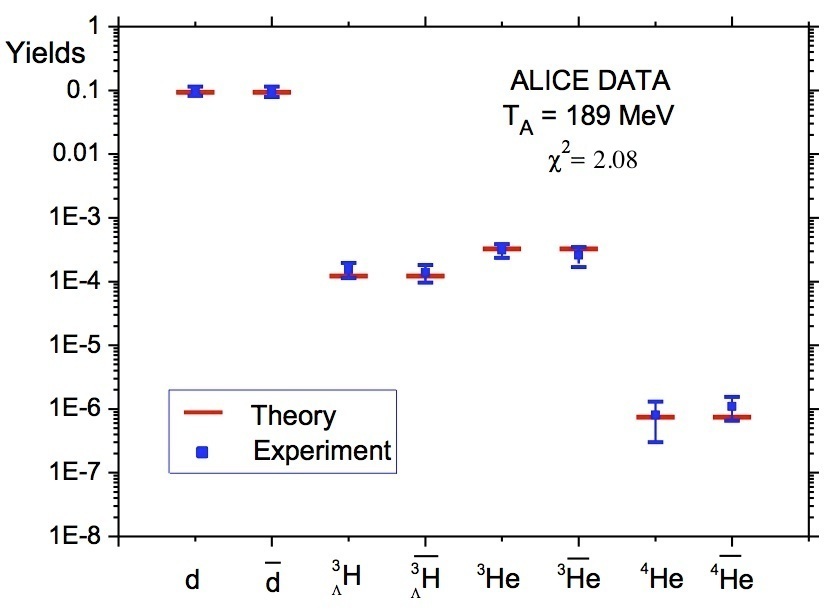}
	}

	\caption{ {\bf Left panel.} The yields of nuclear clusters at the CFT temperature $T=168.5$ MeV which corresponds to   $\min\chi^2_A (V)$ for the EoS (\ref{KAB_Eq21}), (\ref{KAB_Eq22}):  the ALICE data (symbols) vs. theoretical description (bars).
	   {\bf Right panel.} Same as in the left panel, but for the EoS 
(\ref{KAB_Eq19}), (\ref{KAB_Eq20}). $\min\chi^2_A (V)$ corresponds to the CFO temperature  $T=189$ MeV.
}
	\label{KAB_Fig2}
\end{figure}

However, if one analyzes the contributions of $\chi^2_{h}$ and $\chi^2_{A} (V)$ separately (see  Fig. \ref{KAB_Fig1}),
then one finds that their local minima are located at rather different temperatures, but  the sum of $\chi^2_{h}$ and $\chi^2_{A} (V)$ has a minimum which is located more close to the minimum of $\chi^2_{h}$!  
Moreover, from Fig. \ref{KAB_Fig1} one can see that $\chi^2_{A} (V)$ has a deep minimum even for  the vanishing 
size  of nuclear clusters, i.e. an existence of such a minimum is a generic feature
of the advanced versions of HRGM. 
Hence, we concluded that it is worth to  verify the hypothesis of separate CFO of hadrons and nuclear clusters \cite{KAB_Ref3}.  Hereafter such a model is called M2.  
In this case there are three  fitting parameters, namely the CFO temperature of hadrons $T_h$,  the one of nuclear clusters $T_A$ and the CFO volume $V$ of nuclei.  For the EoS (\ref{KAB_Eq21}), (\ref{KAB_Eq22})  one finds that $T_A^{M2R} \simeq 168.5 -2.6+6.7$ MeV
and  $\chi^2_{tot}/dof|_{M2R} \simeq
(9.1+2.2)/(11+8-3)  = 11.3/16 \simeq 0.706$. In other words, compared to the M1R the mean deviation squared per number 
of degrees of freedom of M2R decreased on 50\%!  

A similar hypothesis verified  with the EoS (\ref{KAB_Eq19}), (\ref{KAB_Eq20}) gives us $T_A^{M2V} \simeq 189 -15+38$ MeV
and  $\chi^2_{tot}/dof|_{M2V} \simeq
(9.1+2.08)/(11+8-3)  = 11.18/16 \simeq 0.699$.  The latter value is very close to the value $\chi^2_{tot}/dof|_{M2R}$
and, hence, we do not face the problem of M1. 
Moreover,  as one can see from Fig. \ref{KAB_Fig2} an excellent  fit quality  provided by  two different EoS is very similar.
However, at first glance we face two new problems, namely  that the CFO  temperatures 
 of nuclear clusters differ essentially from each other and that the  CFO temperature $T_A^{M2V}$  is somewhat   larger than the  
cross-over temperature $T_{co} \simeq  147-170$ MeV  predicted by the lattice formulation of QCD at vanishing value of the  baryonic chemical potential \cite{KAB_lqcd1,KAB_lqcd2}. 

Considering the cross-over temperature value one should remember that the lattice formulation of QCD predicts the 
temperature of an infinite system.  In HIC, on contrary, the formed systems  are small and they have boundary with the vacuum. Therefore, it is not surprising that the temperature of small systems created  in HIC may be higher than the ones predicted by the lattice formulation of QCD.  The validity of this statement  can be seen  from the values of  CFO temperature  found for the RHIC  highest collision energies which are  about 170-175 MeV \cite{IST2,KAB_Chatter15,KAB_pbm17,KAB_Jean}. 

The problem with different values of  CFO temperatures found for M2R and M2V can also be  resolved easily.
Indeed, comparing the left and right panels of   Fig. \ref{KAB_Fig1} one can see that there is a  region of  CFO temperatures  at which the shallow minima of $\chi^2_{A}|_{M2R}$ and $\chi^2_{A}|_{M2V}$ obeying  the following inequalities
 $\chi^2_{A}|_{M2R} \le \min(\chi^2_{A}|_{M2R})+1$  and $\chi^2_{A}|_{M2V}\le \min(\chi^2_{A}|_{M2V})+1$ do overlap. 
 Note that exactly these inequalities are used to determine the most probable range of  CFO temperature. Hence there is 
 a narrow region of  CFO temperatures $T_A^{M2} \in [174; 175.2]$ MeV where both EoS discussed above  provide 
a  fit  of ALICE data of  very similar quality.  
One can estimate the total quality of the fit for $T_A^{M2} \in [174; 175.2]$ MeV and get 
$\chi^2_{tot}/dof|_{M2} \simeq
(9.1+3.2)/(11+8-3)  = 12.3/16 \simeq 0.769$ which is only about 10\% larger than the best fit of the M2 found above. 
Thus, both EoS describing the mixture of hadrons and nuclear clusters agree well with each other for the CFO temperature 
of nuclear clusters 
$T_A^{M2} \simeq  174.6 \pm 0.6$ MeV. This CFO temperature is in  remarkable agreement with the CFO temperature of hadrons found for the highest RHIC energies \cite{IST2,KAB_Chatter15,KAB_pbm17,KAB_Jean}. 

\section{Conclusions}

In this work we found the classical excluded volumes of roomy clusters which can be used to describe the 
yields of  such nuclear clusters as  deuterons,  helium-3,  helium-4  and hyper-triton  
and their antiparticles. From the obtained expression  (\ref{KAB_Eq7})  we deduced  an approximate formula  (\ref{KAB_Eq9}) for the effective hard-core radius 
of nuclear  clusters 
for  a pion dominated medium. The main advantages of  the formula (\ref{KAB_Eq9}) are its simplicity and  that 
it does not require an essential modification of  the HRGM based on the induced surface tension concept (R-approach)  developed in Refs. \citen{IST1,IST2}. However, to verify the validity of such an approach we heuristically derived the novel HRGM with induced surface tension (V-approach)  which  employes the true classical excluded volumes of roomy clusters.  
By construction the both approaches should provide similar results for  a pion dominated medium. 

Using these two approaches  we  analyzed the ALICE data  \cite{KAB_Ref1a,KAB_Ref1b,KAB_Ref1c} on the yields of  nuclear clusters. Our first analysis  (the model M1)  is based on the traditional assumption that the CFO of hadrons and nuclear clusters occurs at the same hyper-surface and, therefore,  the CFO temperature  and volume are the fitting parameters of such a model. Although the found CFO temeperatures of R- and V-aproaches, i.e. the models M1R and M1V,
agree well with each other, we got a paradoxical  result, namely that the quality of the  description of approximate approach
$\chi^2_{tot}/dof|_{M1R} \simeq 1.476$ is essentially higher than the one $\chi^2_{tot}/dof|_{M1V} \simeq 1.83$
obtained by a  more elaborate V-approach.  Thus,  we faced an alternative: (1) either to claim that the nuclear clusters cannot be accurately described by the advanced V-approach  and, hence, the description  of  the nuclear clusters obtained in the oversimplified versions of the HRGM \cite{KAB_Ref2, KAB_Jean, PBM19} is just a kind of  illusion, or (2) to abandon the traditional assumption of a single CFO of hadrons and nuclear clusters. 

A close inspection of the $\chi^2_{A} (V(T))$ as a function of CFO temperature $T$ showed us that $\chi^2_{A} (V(T))$ of  nuclear clusters  has a deep minimum located at higher  temperature  than  the CFO one of M1.  Hence we verified the model M2 in which the CFO of nuclear clusters occurs separately  from CFO of hadrons.  As it was argued a long time ago 
\cite{KAB_EarlyFO1,KAB_EarlyFO2},  at  temperatures above the pion mass one can naturally expect an early  and simultaneous chemical and kinetic freeze-out 
 for heavy particles, which do not produce the resonances with pions. In Refs. \citen{KAB_EarlyFO1,KAB_EarlyFO2} the validity of 
such a hypothesis was demonstrated for  $\Omega$ hyperons and $J/\psi$ and $\psi^{\prime}$ mesons. Later on 
it was shown \cite{SFO, SFO_ind,SFO_ind2,SFO2}  that at the collision energies above SPS ones the CFO of strange hadrons  occurs separately at  temperatures higher than for  the non-strange ones. Thus, our M2  is a natural extension  of  an early CFO 
hypothesis. Moreover, intuitively it is clear that the compact objects as hadrons and  the extended ones as the nuclear clusters may require different conditions of formation and, hence, our M2 is in line  with this expectation. 

In the M2 there appears an additional fitting parameter compared to the M1, 
i.e. the CFO temperature of nuclear clusters $T_A$.
Then we got an opposite situation than for the M1 that 
$\chi^2_{tot}/dof|_{M2R} \simeq 0.706$ and 
$\chi^2_{tot}/dof|_{M2V}  \simeq 0.699$ of R- and V-approaches agree well with each other, but their CFO temperatures are rather different.
As one can see from  Fig. \ref{KAB_Fig2} both versions  of  the M2 provide an unprecedentedly accurate description 
of nuclear clusters`  yields. This fact motivated us to resolve the problem with different values of CFO temperature of nuclear clusters. It turns out that within   a narrow region of  CFO temperatures $T_A^{M2} = 174-175.2$ MeV the both approaches
agree very  well with each other, as it was expected in the first place. Moreover, we found that in this narrow region of  CFO temperatures the description of both approaches is still unprecedentedly accurate with $\chi^2_{tot}/dof|_{M2} \simeq 0.769$.
These findings are of principal importance for the  HIC phenomenology, since they may shed the light on the mechanism of early freeze-out, both the chemical and the kinetic one,  of  nuclear clusters and shake  the validity of fits obtained with oversimplified versions of the HRGM.

\vspace*{4.4mm}
{\small 
{\bf Acknowledgements.}
The authors are thankful to O. V. Vitiuk and E. S. Zherebtsova for fruitful discussions and helpful comments. 
K.A.B., B.E.G., V.V.S., O.I.I.  and G.M.Z. 
acknowledge the partial support by the Program
of Fundamental Research in High Energy and Nuclear Physics launched 
by the Section of Nuclear Physics of the NAS
of Ukraine. 
 V.V.S. and  O.I.I. are thankful for the support by the Funda\c c\~ao para a Ci\^encia e Tecnologia (FCT), Portugal, under the project UID/FIS/04564/2019. The work of O.I.I. was supported by  the project CENTRO-01-0145-FEDER-000014 through CENTRO2020 program, and  POCI-01-0145-FEDER-029912 with financial support from POCI, in its FEDER component, and by the FCT/MCTES budget through national funds (OE).
The work of L.V.B. and E.E.Z. was supported by the Norwegian 
Research Council (NFR) under grant No. 255253/F50 - CERN Heavy Ion 
Theory. 
L.V.B. and K.A.B. thank the Norwegian Agency for International Cooperation and Quality Enhancement in Higher Education for financial support, grant 150400-212051-120000 "CPEA-LT-2016/10094. 
A.V.T. acknowledges partial support from RFBR under grant No.
18-02-40086 and from the Ministry of Science and Higher Education of the Russian Federation
Project No 0723-2020-0041.
D.B.B. received funding from the RFBR under  grant No. 18-02-40137.
The authors are thankful  to the  COST Action CA15213 for supporting their networking. 
}

\end{document}